\documentclass[preprint,floatfix,prb,showpacs,color,superscriptaddress,epsfig]{revtex4}
\usepackage{graphicx}
\usepackage{amsfonts}
\usepackage{amsmath}
\usepackage{amssymb}
\usepackage{dcolumn}
\usepackage{multirow}
\usepackage{epsfig}
\begin{document}
\title{Chemical trend of exchange couplings in diluted magnetic II-VI semiconductors}
\author{T. Chanier}
\affiliation{Institut de Mat\'eriaux, Micro\'electronique et
Nanosciences de Provence, Facult\'e St.\ J\'er\^ome, Case 142,
F-13397 Marseille Cedex 20, France}
\author{F. Virot}
\affiliation{Institut de Mat\'eriaux, Micro\'electronique et
Nanosciences de Provence, Facult\'e St.\ J\'er\^ome, Case 142,
F-13397 Marseille Cedex 20, France}
\author{R. Hayn}
\affiliation{Institut de Mat\'eriaux, Micro\'electronique et
Nanosciences de Provence, Facult\'e St.\ J\'er\^ome, Case 142,
F-13397 Marseille Cedex 20, France}
\date{\today}

\begin{abstract}
We have calculated the chemical trend of magnetic exchange
parameters ($J_{dd}$, $N \alpha$, and $N \beta$) of Zn-based II-VI
semiconductors ZnA (A=O, S, Se, and Te) doped with Co or Mn. We show
that a proper treatment of electron correlations by the LSDA+$U$
method leads to good agreement between experimental and theoretical
values of the nearest-neighbor exchange coupling $J_{dd}$ between
localized 3$d$ spins in contrast to the LSDA method. The exchange
couplings between localized spins and doped electrons in the
conduction band $N \alpha$ are in good agreement with experiment as
well. But the values for $N \beta$ (coupling to doped holes in the
valence band) indicate a cross-over from weak coupling (for A=Te and
Se) to strong coupling (for A=O) and a localized hole state in
ZnO:Mn. That hole localization explains the apparent discrepancy
between photoemission and magneto-optical data for ZnO:Mn.
\end{abstract}



\maketitle

\section{Introduction}

After the seminal discovery of ferromagnetism in GaAs:Mn
\cite{Matsukura98} with a critical temperature $T_c$ as high as 110
K there is worldwide a renewed interest in diluted magnetic
semiconductors (DMS). Recently, the Curie temperature in GaAs:Mn
could be pushed to values of about 180 K  by a careful control of
the annealing conditions during the growth process. \cite{Olejnik08}
There is a great search activity to look for alternative materials,
especially in the class of II-VI semiconductors (SC). Ferromagnetism
(FM) in diluted II-VI SC is known for a long time with up to now low
$T_c$ values, however. \cite{Cochrane74} They also serve as model
materials since they allow to control the magnetic ions and the
doped charge carriers independently. In such a way it was possible
to demonstrate the carrier-induced mechanism of the ferromagnetic
state in Pb-doped SnTe:Mn \cite{Story86} or in $p$-doped ZnTe:Mn.
\cite{Ferrand01}

The DMS combine ferromagnetism with the conductivity properties of
semiconductors. Therefore, they are ideal materials for applications
in spintronics where not only the electron charge but also the spin
of the charge carrier is used for information processing. For
instance, they allow to resolve the conductivity mismatch problem
which hinders a high polarizability of injected electrons in a
ferromagnetic metal/semiconductor junction. \cite{Schmidt00}

The ferromagnetism in the traditionally known DMS arises due to
Zener's $p$-$d$ exchange mechanism. \cite{Zener50} The 3$d$
transition metal impurities lead to localized spins ${\bf S}_i$.
Hole doping into the valence band (either by the 3$d$ transition
metals itself or by other acceptor impurities) provides charge
carries whose spins interact with the 3$d$ spins. This local $p$-$d$
exchange coupling $J_{pd}^v=N \beta$ leads to a parallel
arrangements of the magnetic moments since a ferromagnetic state
allows a higher mobility of the doped holes. For a high doping level
the material becomes more metallic and the mechanism changes to a
RKKY-like interaction.

From this argumentation follows immediately that the crucial
parameter to increase $T_c$ is the $J_{pd}^v$ coupling. Indeed, a
simple theory of Zener's $p$-$d$ exchange mechanism \cite{Dietl01}
gives $T_c \propto (J_{pd}^v)^2 x_h$ where $x_h$ is the hole doping
level. It can be expected from general grounds that a decreasing
anion-cation distance leads to an increase of the $p$-$d$
tight-binding hopping parameter $t_{pd}$, and consequently to an
increase of $J_{pd}^v$. That reasoning lead Dietl {\em et al}
\cite{Dietl01} to the proposal of room temperature ferromagnetism in
Mn-doped ZnO or GaN, respectively, which created a tremendous
activity and numerous reports on room temperature FM in II-VI DMS or
similar materials. \cite{Lee02,Prellier03,Sharma03}

However, there are serious doubts whether the reported room
temperature ferromagnetism belongs really to the same class of
ferromagnetism as that one observed in GaAs:Mn or ZnTe:Mn which is
based on Zener's $p$-$d$ exchange mechanism. For instance, in ZnO:Co
ferromagnetism was reported in samples produced by laser
ablation,\cite{Prellier03,Sharma03} or by the sol-gel
method,\cite{Lee02} whereas other samples fabricated by precursor
deposition,\cite{Lawes05} or molecular beam epitaxy (MBE)
\cite{Sati06,Yoon03} showed no signs of ferromagnetism and
antiferromagnetic couplings between nearest neighbor $3d$ spins. It
is highly probable that the observed ferromagnetic effects in ZnO:Co
are due to uncompensated spins at the surface of Co-rich
antiferromagnetic nanoclusters. \cite{Dietl07}

The proposal of Dietl {\em et al} \cite{Dietl01} was based on simple
model calculations and qualitative arguments. There is a real need
for a parameter free {\em ab-initio} study of the relevant exchange
parameters in II-VI semiconductors to put the expected chemical
tendency on a firm basis. Such a calculation of the nearest neighbor
couplings of local spins $J_{dd}$ and the $p$-$d$ exchange couplings
$J_{pd}^v$ and $J_{pd}^c$ with valence and conduction bands,
respectively, is presented here. We considered the series of Co- and
Mn-doped ZnA with the anions A=Te, Se, S and O.

To achieve our goal we had to solve two theoretical problems. First
of all, the local spin density approximation (LSDA) is not
sufficient. It leads to wrong predictions of FM in ZnO:Co even
without additional hole doping, \cite{Sato01} to too large values of
$|J_{dd}|$ for ZnO:Mn, and to the wrong (FM) sign of one of the two
nearest neighbor exchange couplings in wurtzite ZnO:Co.
\cite{Chanier06} It was shown that this deficiency of LSDA can be
repaired by taking into account the strong Coulomb correlation in
the 3$d$ shell by the LSDA+$U$ method. To choose the $U$ values we
have to take into account that they decrease in the series from O to
Te due to an increase of screening effects. The values of $J_{dd}$
are very well known experimentally in this series. Therefore they
can be used to check the chosen $U$ values. We will show below that
for reasonable values of $U$ we obtain $J_{dd}$ in good agreement
with experimental results and we may explain the chemical tendency.

The second theoretical problem concerns the $p$-$d$ exchange
coupling between the localized spins and the holes in the valence
band $J_{pd}^v$. This coupling leads to the giant Zeemann effect
\cite{Furdyna88} and it is seen in our calculations by a band-offset
$\Delta E^v$ between spin up and spin down of the valence band. For
small values of $J_{pd}^v$ (which means also small values of
$t_{pd}$) both splittings, the experimental and the theoretical one,
are proportional to the magnetic impurity concentration $x$. In that
weak coupling regime the $p$-$d$ coupling can be simply calculated
by using the proportionality between splitting and $x$. We will
show, however, that there are more and more deviations from $\Delta
E^v \propto x$ if we go from ZnTe to ZnO. The exchange values
obtained in that manner seem to depend on the concentration of
magnetic impurities. We solve that problem by a fit to the
Wigner-Seitz approach of Benoit a la Guillaume {\em et al}.
\cite{Benoit92} Our results prove that we reach the strong coupling
limit for ZnO. As we will show below, in that case the impurity
potential is so strong that it can bind a hole for ZnO:Mn, whereas
ZnO:Co is close to the localized limit.

Our ab-initio results strengthen the recent model calculations in
Ref.\ \onlinecite{Dietl08}. That work was aimed to explain the
tremendous difference between the experimental $J_{pd}^v$ values
obtained from photoemission and magneto-optics, especially in ZnO
and GaN. \cite{Pacuski08} It was argumented that this difference
arises due to state localization which is confirmed by our ab-inito
calculations below. But we also will show that our results for
$|N\beta|$ are much smaller than those evaluated earlier from
photoemission for ZnO:Mn (-2.7 eV (Ref.\ \onlinecite{Mizokawa02}) or
-3.0 eV (Ref.\ \onlinecite{Okabayashi04})) and ZnO:Co (-3.4 eV
(Ref.\ \onlinecite{Blinowski02})) and which were used as model input
parameters in Ref.\ \onlinecite{Dietl08}.

The organization of our paper is as follows. After presenting the
super-cell method in Sec. II, we discuss the nearest neighbor
exchange coupling in Sec. III. That fixes the $U$ values
unambiguously. In Sec. IV we present our results for
$J_{pd}^v=N\beta$ and $J_{pd}^c=N\alpha$. Finally, in Sec. V we
discuss the arguments in favor of a localized state in ZnO:Mn.

\section{Super-cell calculations}

We used super-cell calculations to determine the exchange couplings
$J_{dd}$, $N \alpha$, and $N \beta$. Since we are mainly interested
in the chemical tendency within the II-VI series we restrict our
study to the zinc-blende structure. All compounds of the series
exist in that modification, even ZnO as epitaxial layer. To
calculate $J_{dd}$ we used super-cells of the form T$_2$Zn$_6$A$_8$
with the transition metals T=Co or Mn and with the anions A=O, S,
Se, and Te. In those super-cells the magnetic ions build chains. The
exchange constants are then determined by comparing the total
energies of ferro- and antiferromagnetic arrangements. We have
checked that the influence of finite size effects is negligible (not
larger than 6 per cent for $J_{dd}$) by performing some calculations
with T$_2$Zn$_{14}$A$_{16}$ super-cells.

For $J_{pd}^v$ we used super-cells with three different
concentrations of magnetic ions, $x=1/4$, $x=1/8$, and $x=1/32$,
i.e. TZn$_3$A$_4$, TZn$_7$A$_8$ and TZn$_{31}$A$_{32}$. As will be
explained below, these numerical results have to be fitted with the
Wigner-Seitz approach to obtain $J_{pd}^v$. In all calculations we
used the experimental lattice constants $a=6.101$ \AA, 5.668 \AA,
and 5.410 \AA \ for ZnTe, ZnSe, and ZnS, respectively.
\cite{Jamieson80,Handbook90} For ZnO we used a lattice constant
$a=4.557$ \AA \ which gives the same unit cell volume as the
experimental value. (Bulk ZnO has $a=3.2427$ \AA \ and $c=5.1948$
\AA \ in the wurtzite structure.) \cite{Sabine69}

The super-cell calculations were performed using the full-potential
local-orbital (FPLO) band structure scheme. \cite{Koepernik99} In
the FPLO method (version FPLO5) a minimum basis approach with
optimized local orbitals is employed, which allows for accurate and
efficient total energy calculations. For the present calculations we
used the following basis set: Zn,Co,Mn: $3s3p$:$4s4p3d$, O:
$2s2p$;$3d$, S: $3s3p3d$, Se: $4s4p3d$, and Te: $5s5p4d$. The
site-centered potentials and densities were expanded in spherical
harmonic contributions up to $l_{max}=12$.

The exchange and correlation potential was treated in two different
ways. First, the local spin-density approximation (LSDA) was used in
the parametrization of Perdew and Wang. \cite{Perdew92} However, as
will be shown below in more detail, this approximation has severe
deficiencies in the present case. The energetical positions of the
Co(Mn) $3d$ states with respect to the valence band are incorrectly
given in the LSDA calculation. They are expected to be much lower in
energy and this correlation effect was taken into account by using
the FPLO implementation of the LSDA+$U$ method in the atomic limit
scheme. \cite{Anisimov91,Eschrig03} The convergence of the total
energies with respect to the ${\bf k}$-space integrations were
checked for each of the super-cells independently. The calculations
for each cell were first performed within the LSDA approximation
using basis optimization. The LSDA+$U$ calculations were then made
starting from the LSDA optimized basis but with no basis
optimization in the self-consistency cycle in order to obtain
convergence. The Slater parameters $F^2$ and $F^4$ for Mn and Co in
the LSDA+$U$ calculations were chosen close to atomic values, namely
$F^2=7.4$ eV and $F^4=4.6$ eV (corresponding to the Hund exchange
coupling $J_H=0.86$ eV) for Mn, and $F^2=7.9$ eV and $F^4=5.0$ eV
($J_H=0.92$ eV) for Co. The Slater parameter $F^0=U$, however, is
much more screened and its influence has been investigated more in
detail (see below).

\begin{figure}
\includegraphics[scale=0.35,angle=0]{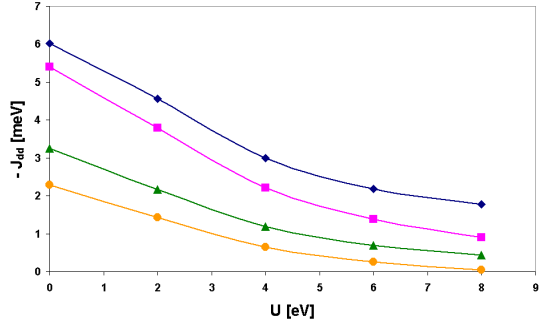}
\caption{(Color online) Calculated exchange couplings $J_{dd}$ for
ZnA:Mn (from above to below: A=0 (blue), S (red), Se (green), and Te
(yellow) as a function of the Coulomb correlation $U$ in the 3$d$
shell.} \label{fig0}
\end{figure}


\begin{figure}
\includegraphics[scale=0.35,angle=0]{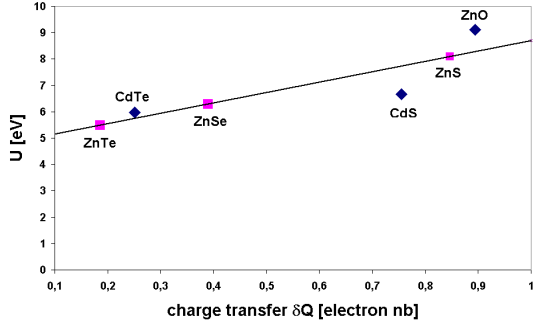}\\
\caption{(Color online) Coulomb correlation $U$ in the 3$d$ shell of
Mn-impurities in II-VI-SC as function of charge transfer $\delta Q$.
The values for CdTe, CdS, and ZnO are taken from Ref.\
\onlinecite{Gunnarson89} (constrained DFT-calculations, blue
diamonds) and the other values (red squares) by linear interpolation
corresponding to the calculated charge transfer.} \label{fig1}
\end{figure}

\section{d-d Exchange couplings}

\begin{table}[h!]
\caption{\label{table1a}Comparison of calculated and experimental
values for the nearest neighbor exchange couplings $J_{dd}$ for
ZnA:Mn.}
\begin{ruledtabular}
\begin{tabular}{cccccccc}
 &  \multicolumn{1}{c}{exp.} & \multicolumn{2}{c}{LSDA+$U$} & LSDA
\\  & $J_{dd}^{exp}$(meV)  & $J_{dd}$(meV) & $U$(eV)
& $J_{dd}$(meV)
  \\\hline
ZnO:Mn & -2.09\footnotemark[1] & -2.18 & 6 & -6.02 \\
ZnS:Mn & -1.41\footnotemark[2] ; -1.39\footnotemark[3]
& -1.39 & 6  & -5.41 \\
ZnSe:Mn &  -1.05\footnotemark[4] ; -1.06\footnotemark[3]
& -1.19 & 4  & -3.25 \\
ZnTe:Mn & -0.75\footnotemark[5] ; -0.76\footnotemark[6]
& -0.65 & 4  & -2.29 \\
\end{tabular}
\end{ruledtabular}
\footnotetext[1]{Magnetization step method,
Ref.~\onlinecite{Gratens04}, first neighbor in the (a,b) plane of
the wurtzite structure.} \footnotetext[2]{Magnetization step method,
Ref.~\onlinecite{Shapira02}.} \footnotetext[3]{Inelastic neutron
scattering, Ref.~\onlinecite{Giebultowicz87}.}
\footnotetext[4]{Magnetization step method,
Ref.~\onlinecite{Foner89}.} \footnotetext[5]{Magnetization step
method, Ref.~\onlinecite{Lascaray87}.} \footnotetext[6]{Inelastic
neutron scattering, Ref.~\onlinecite{Corliss86}.}

\end{table}
\begin{table}
\caption{\label{table1b}Comparison of calculated and experimental
values for the nearest neighbor exchange couplings $J_{dd}$ for
ZnA:Co. }
\begin{ruledtabular}
\begin{tabular}{ccccccccc}
 & \multicolumn{1}{c}{exp.} & \multicolumn{2}{c}{LSDA+$U$} & LSDA
\\  & $J_{dd}^{exp}$(meV)  & $J_{dd}$(meV) &
$U$(eV) & $J_{dd}$(meV)
  \\\hline
ZnO:Co & -2.0\footnotemark[1] & -1.73 & 6 & -1.39 \\
ZnS:Co & -4.09\footnotemark[2] & -4.13 & 4  & -7.26 \\
ZnSe:Co & -4.26\footnotemark[3] & -3.36 & 4  & -6.26 \\
ZnTe:Co & -3.27\footnotemark[3]  & -3.32 & 4  & -6.94 \\
\end{tabular}
\end{ruledtabular}
\footnotetext[1]{Inelastic neutron scattering,
Ref.~\onlinecite{Stepanov07}, first neighbor in the (a,b) plane of
the wurtzite structure.} \footnotetext[2]{Inelastic neutron
scattering, Ref.~\onlinecite{Giebultowicz90}.}
\footnotetext[3]{Inelastic neutron scattering,
Ref.~\onlinecite{Giebultowicz90b}.}

\end{table}

In this Section we are going to determine the exchange couplings
between two localized magnetic ions. We are considering two nearest
neighbor impurities, each carrying a local spin ${\bf S}_i$. Then,
the Heisenberg Hamiltonian for a localized pair of spins is given by
\begin{equation}
H=-2J_{dd} {\bf S}_i {\bf S}_j \; .
\end{equation}
The corresponding total energies per magnetic ion for ferromagnetic
(FM) and antiferromagnetic (AFM) arrangements of the two spins,
$E_{FM}$ and $E_{AFM}$, lead to the energy difference between the FM
and AFM states:
\begin{equation}
\Delta E = \frac{E_{FM}-E_{AFM}}{2} = -\frac{J_{dd}}{2}S_T(S_T+1) \;
, \label{pair}
\end{equation}
where $S_T$ is the total spin of two parallel spins $S$, i.e.\
$S_T=3$ or 5 for Co or Mn. That energy difference can be compared
with the corresponding energy differences of isolated pairs in the
large super-cells. Those super-cells where the magnetic ions form
chains are different, however. Then, each magnetic ion has two
nearest neighbor magnetic ions which doubles approximatively the
previous energy difference (\ref{pair}). The exact energy difference
between FM and AFM states of a Heisenberg chain is slightly
different, but that is unimportant for our present argumentation.

The calculated exchange constants $J_{dd}$ show a strong variation
with $U$. That is illustrated in Fig. \ref{fig0} for the Mn-doped
compounds. In that case the tendency is monotonous, i.e. the
increase of $U$ leads to a decrease of $J_{dd}$. A similar tendency
is visible for ZnA:Co with the exception of ZnO:Co where the LSDA
exchange constant is only -1.39 meV and not of the order of -6...-7
meV like for the other compounds. This exception is due to
ferromagnetic contributions in ZnO:Co as analyzed in Ref.\
\onlinecite{Chanier06}.

The experimental values of $J_{dd}$ are known with great accuracy by
magnetization step measurements or inelastic neutron scattering (see
Tables \ref{table1a}, \ref{table1b}). The comparison of experimental
and theoretical values shows that the LSDA method strongly
overestimates the exchange couplings. In our method the Hubbard
correlation has to be chosen between 4 and 6 eV to obtain the
correct exchange couplings. The precise value of $U$ has also a
chemical tendency. That was revealed in Ref.\
\onlinecite{Gunnarson89} and can be explained since the compounds
ZnA become less and less ionic in going from A=Zn to A=Te. The
decrease of ionicity can be measured by a decrease of the charge
transfer towards the magnetic ion in the series (Fig.\ \ref{fig1}).
The charge transfer is correlated with the calculated $U$ value in
the constrained density functional calculation. \cite{Gunnarson89}
Taking into account this chemical tendency we chose the $U$ values
of Tables \ref{table1a}, \ref{table1b} to calculate $J_{dd}$ (and $N
\alpha$, $N \beta$ in the next Chapter). Those values for $U$ are
slightly smaller than that one calculated in Ref.\
\onlinecite{Gunnarson89} since the FPLO and LMTO (linearized muffin
tin orbitals) implementations of the LSDA+$U$ method are not
equivalent. Taking into account the restricted accuracy of our
procedure we varied $U$ in steps of 2eV. Then we obtain the
theoretical results of Tables \ref{table1a}, \ref{table1b} which are
in good agreement with the experimental values.

\section{p-d exchange couplings}

The localized magnetic moments ${\bf S}_i$ which are provided by the
magnetic ions Co$^{2+}$ or Mn$^{2+}$ interact with the spin of doped
holes ${\bf s}$. This interaction can be parametrized in the
continuum approximation in the form:
\begin{equation}
\hat{H}= - \beta \sum_i {\bf S}_i {\bf s} \delta ({\bf r} - {\bf
R}_i) \; , \label{eq41}
\end{equation}
where the magnetic impurities are placed at ${\bf R}_i$. A similar
interaction exists with the spin of doped electrons which is usually
denoted by the parameter $\alpha$. If we transform the Hamiltonian
into a lattice model, the interaction (\ref{eq41}) becomes
\begin{equation}
\hat{H}= - J_{pd}^v \sum_i {\bf S}_i {\bf s}_i  \; , \label{eq42}
\end{equation}
with the sum over all lattice sites $i$ which are occupied by
magnetic impurities, and where ${\bf s}_i$ is the local spin
operator of the doped hole in the lattice representation. Both
parameters are connected by $J_{pd}^v=N\beta$ where $N$ is the
number of cations per volume ($N=4/a^3$ in the zinc blende
structure). One possibility to measure $N\beta$ is photoemission
where the hole in the valence band is created during the
photoemission process. Another possibility is magnetooptics which
measures the giant Zeeman effect of excitons, i.e.\ electron-hole
pairs.

We calculated the $p$-$d$ exchange coupling with super-cells having
impurity concentrations of $x=1/4$, 1/8, and 1/32 magnetic ions. The
$p$-$d$ exchange coupling leads to a valence band and conduction
band offset between spin up and spin down $\Delta E^v$ and $\Delta
E^c$. In the case of weak $p$-$d$ coupling, this band offset is
proportional to the impurity concentration $x$, i.e.\ it can be
calculated in mean-field theory. That can be clearly observed in our
numerical data and the corresponding exchange couplings are then
simply given by
\begin{equation}
J_{pd}^c=N\alpha=\frac{\Delta E^c}{x\langle S \rangle } \;
\mbox{and} \; J_{pd}^v=N\beta=\frac{\Delta E^v}{x\langle S \rangle }
\; , \label{eq43}
\end{equation}
where $\langle S \rangle = \langle M \rangle /(2 \mu_B )$ is the
mean value of the local spin calculated within the ab-initio
approach. For the Mn compounds, the calculated magnetization values
are very close to saturation ($\langle M \rangle / \mu_B$ = 5.00,
4.85, 4.83, and 4.85 in the series with the anions A=O, S, Se, and
Te, respectively) but there are stronger deviations from the local
value $S=3/2$ for the Co ones ($\langle M \rangle / \mu_B$ = 3.00,
2.65, 2.75, and 2.61 in the same series).

The mean-field approach works very well for $N\alpha$ which has
small values in all cases. The reason is the small coupling between
the conduction band, which is built by Zn $4s$-$4p$ hybridized
states, with the impurity states. The calculated values are also in
excellent agreement with the available experimental data (see Tables
\ref{table2a} and \ref{table2b}).

\begin{figure}
\includegraphics[scale=0.35,angle=-90]{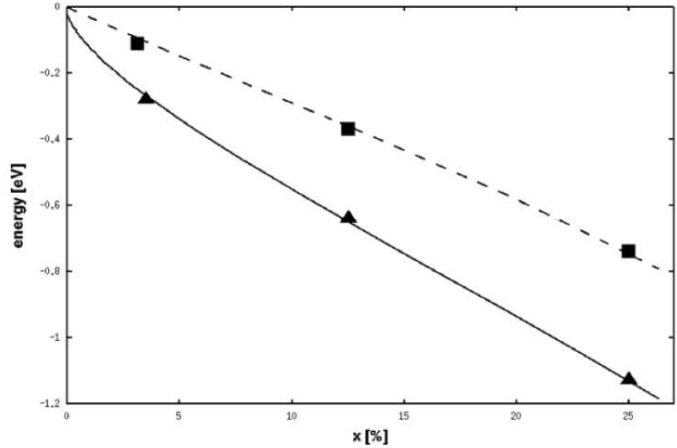}
\caption{Valence band-edge spin splitting $\Delta E^v$ of ZnO:Mn
(triangles) and ZnTe:Mn (squares) calculated with the LSDA+$U$
method. The solid (dashed) lines represent the fit to the
Wigner-Seitz model.} \label{figs}
\end{figure}

The situation is different for $N\beta$. The valence band is built
by the anion $p$-orbitals which have generally a large overlap with
impurity states. Therefore, $N\beta$ is much more important than
$N\alpha$. And this tendency is increased when the lattice constant
diminishes in going from Te to O. As a consequence, the mean-field
description, and the proportionality between band-offset and
impurity concentration breaks down. Historically, the deviation from
the mean-field picture was first observed experimentally for CdS:Mn.
\cite{Ryabchenko87} In our calculations, deviations from the
mean-field behavior are especially visible for doped ZnO and ZnS.
They are mostly pronounced for ZnO:Mn (see Fig.\ \ref{figs}) where a
localized state appears which means that $\Delta E^v$ tends to a
constant value for $x \to 0$ (see next Section). Formally, the
mean-field calculation of $N\beta$ (\ref{eq43}) leads then to a
divergent value which illustrates the discussed break-down in the
most prominent way. This can also be interpreted as a crossover from
the weak coupling to the strong coupling regime in the series going
from Te to O. Since the localization is expected to disappear for
higher impurity concentrations (visible in a band merging of the
localized state with the valence band in the density of states)
there is some justification to use the mean-field formula
(\ref{eq43}) for $x=1/4$. The values of $N \beta^{MF}$ calculated in
that way are displayed in Tables \ref{table2a} and \ref{table2b}.

To resolve the deviations from the mean-field behavior a
Wigner-Seitz approach was developed. \cite{Benoit92} We will use it
to calculate $N\beta$ more accurately (see also Ref.\
\onlinecite{Sanvito01} for GaAs:Mn). In that theory, the valence
band is described in the effective mass approximation with a spin
dependent impurity potential. The Hamiltonian for one impurity has
the form:
\begin{equation}
\hat{H}= - \frac{\hbar^2}{2m^*} \frac{\partial^2}{\partial {\bf
r}^2} + ( W - J_{pd}^v {\bf S}_i {\bf s}) \Theta(b-r) \; .
\label{eq44}
\end{equation}
Replacing the spin operator ${\bf s}$ by $s_z$ we obtain a spin
polarized scattering potential
\begin{equation}
\hat{H}_{\sigma}= - \frac{\hbar^2}{2m^*} \frac{\partial^2}{\partial
{\bf r}^2} + U_{\sigma} \Theta(b-r) \; , \label{eq45}
\end{equation}
where $\sigma=+1 (-1) = \uparrow (\downarrow)$ and
$U_{\sigma}=W-\sigma S J_{pd}^v / 2$ with the local spin $S=5/2$ and
3/2 for Mn and Co, respectively. The muffin-tin radius of the
scattering potential $b$ was fixed such that the corresponding
spheres around the cations fill in completely the space of the
solid, i.e.\ $(4 \pi/3) b^3 = a^3/4$. The finite concentration of
impurities is taken into account by the condition that the
derivative of the wave function $\Psi^{\prime} (R) =0$ vanishes at
the mean radius $R$ around each impurity which is determined by $(4
\pi /3) R^3=1/(Nx)$. The scattering problem for each spin direction
is easy to solve \cite{Benoit92} and the lowest eigenvalue for spin
up (down) $E_{\uparrow}$ ($E_{\downarrow}$) is given by a
transcendental equation. The valence band splitting can be expressed
as
\begin{equation}
\Delta E^v=E_{\downarrow}-E_{\uparrow} =\frac{4\pi b^3}{3} N x
\left[ U_{\downarrow} \delta(x,\eta_{\downarrow})-U_{\uparrow}
\delta(x,\eta_{\uparrow}) \right] \; , \label{46}
\end{equation}
where $\delta(x,\eta_{\sigma})=|E_{\sigma}/E_{\sigma}^{MF}|$ is the
ratio of this eigenvalue to the mean-field result
\begin{equation}
E_{\sigma}^{MF}=\frac{4}{3} \pi b^3 U_{\sigma} N x \; . \label{47}
\end{equation}
The deviation is controlled by the dimensionless fitting parameter
\begin{equation}
\eta_{\sigma}=\frac{U_{\sigma}}{|U_c|}=2m^*U_{\sigma}\left(\frac{2b}{\pi
\hbar}\right)^2  \; , \label{48}
\end{equation}
where $U_c$ is the critical potential value for the bound state
creation.

\begin{table*}
\caption{\label{table2a}Comparison of experimental and theoretical
values of the $p$-$d$ exchange couplings with the conduction
($N\alpha$) and valence ($N\beta$) bands for ZnA:Mn (A=O, S, Se, and
Te). The theoretical results were obtained by the LSDA+$U$ method,
analyzed within the mean-field approximation ($N\alpha^{MF}$ and
$N\beta^{MF}$) and the Wigner-Seitz approach \cite{Benoit92}
($N\beta^{WS}$). Also given are the dimensionless coupling
parameters $\eta_{\uparrow}$ and $\eta_{\downarrow}$ of the
Wigner-Seitz approach.}
\begin{ruledtabular}
\begin{tabular}{cccccccc}
  & $N\alpha^{exp}$(eV) & $ N\beta^{exp}$(eV) & $N \alpha^{MF}$(eV) &
  $N\beta^{MF}$(eV) & $N\beta^{WS}$(eV)  & $\eta_{\uparrow}$ &
  $\eta_{\downarrow}$
\\ \hline
ZnO:Mn & --- & $|0.1|$\footnotemark[1] ; -2.7\footnotemark[2] ;
-3.0\footnotemark[3]& 0.38 & -1.81 &
-1.42 & 0.08 & -1.12 \\
ZnS:Mn & --- & -1.3\footnotemark[2] & 0.11 & -1.39
& -1.12 & 0.07 & -0.36 \\
ZnSe:Mn & 0.26\footnotemark[4] & -1.31\footnotemark[4] ;
-1.0\footnotemark[2] & 0.29 & -1.46 &
 -1.23 & 0.05 & -0.29 \\
ZnTe:Mn & 0.18\footnotemark[5]  & -1.05\footnotemark[5] ;
-0.9\footnotemark[2] & 0.23 & -1.22 &
 -1.02 & 0.04 & -0.29 \\
\end{tabular}
\end{ruledtabular}
\footnotetext[1]{Magneto optical measurements,
Ref.~\onlinecite{Przezdziecka06}.} \footnotetext[2]{Photoemission
spectroscopy, Ref.~\onlinecite{Mizokawa02}.}
\footnotetext[3]{Photoemission spectroscopy,
Ref.~\onlinecite{Okabayashi04}.} \footnotetext[4]{Magneto optical
measurements, Ref.~\onlinecite{Twardowski83}.}
\footnotetext[5]{Magneto optical measurements,
Ref.~\onlinecite{Heimann84}.}
\end{table*}

\begin{table*}
\caption{\label{table2b}The same as Table \ref{table2a} but for
ZnA:Co.}
\begin{ruledtabular}
\begin{tabular}{cccccccc}
  & $N\alpha^{exp}$(eV) & $ N\beta^{exp}$(eV) & $N \alpha^{MF}$(eV) &
  $N\beta^{MF}$(eV) & $N\beta^{WS}$(eV)  & $\eta_{\uparrow}$ &
  $\eta_{\downarrow}$
\\ \hline
ZnO:Co & --- & 1.0 (or -0.6)\footnotemark[1] ; -3.4\footnotemark[2]
& 0.34 & -1.82 & -1.36 &
0.34 & -0.36 \\
ZnS:Co & --- & --- & 0.21 & -2.64 & -2.24 & 0.03 & -0.49 \\
ZnSe:Co & ---  & -2.2\footnotemark[3] & 0.33 & -2.50 & -1.98 & 0.02 & -0.31 \\
ZnTe:Co & 0.31\footnotemark[4] & -3.03\footnotemark[4] & 0.28 &
-2.44 & -1.88 & 0.04 & -0.34 \\
\end{tabular}
\end{ruledtabular}
\footnotetext[1]{Magneto optical measurements,
Ref.~\onlinecite{Pacuski06}.} \footnotetext[2]{X-ray absorption,
Ref.~\onlinecite{Blinowski02}.} \footnotetext[3]{Magneto optical
measurements, Ref.~\onlinecite{Twardowski93}.}
\footnotetext[4]{Magneto optical measurements,
Ref.~\onlinecite{Zielinski96}.}
\end{table*}

For the fit we  used the gap and the band-offset calculated within
the LSDA+$U$ approach for the three concentrations mentioned above.
We used the experimental values for the effective masses
$m^*/m=0.22$, 0.21, 0.32, and 1.0 for ZnA, A=Te, Se, S, and O,
respectively. These values were obtained by averaging over the
transversal (light) and longitudinal (heavy) effective masses
according to $3/m^*=2/m_t+1/m_l$. \cite{Benoit92} In Fig.\
\ref{figs} we compare the weak coupling case (represented by
ZnTe:Mn) having a linear dependence of the band-offset on the
impurity concentration $x$ with the strong coupling compound ZnO:Mn
showing clear deviations from linearity. The Wigner-Seitz approach
fits well our numerical data and leads to a localized state for
ZnO:Mn. A summary of all the results is presented in Tables
\ref{table2a} and \ref{table2b}.

\section{Localized state}

\begin{figure}
\includegraphics[scale=0.30,angle=-90]{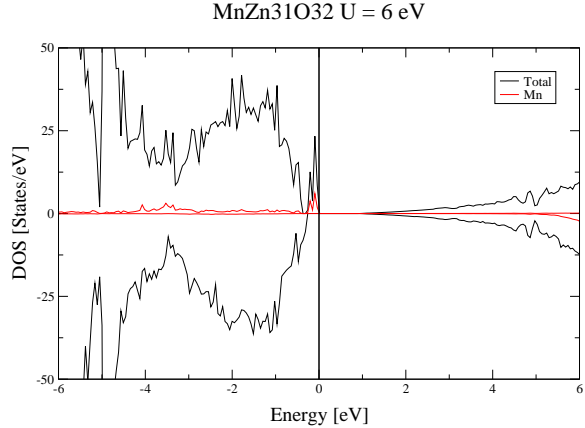}\hskip1cm
\caption{(Color online) LSDA+$U$ density of states for
MnZn$_{31}$O$_{32}$ ($U=6$ eV, black: total DOS, red: partial
Mn-DOS).}\label{fig2}
\end{figure}

The Wigner-Seitz fit for ZnO:Mn results in the dimensionless
coupling parameter $\eta_{\downarrow}=-1.12$ corresponding to a
localized hole state. That is also directly visible in the density
of states (DOS) of MnZn$_{31}$O$_{32}$ (see Fig.\ \ref{fig2}). A
split band appears for $x=1/32$, but not for $x=1/4$. The split band
indicates localization of the hole state, whereas its merging with
the valence band for $x=1/4$ corresponds to a
localization-delocalization transition with increasing doping. (The
accurate description of this transition requires however a better
treatment of disorder and correlation effects.) The Mn $3d$ majority
spin states (upper part of the Figure) are strongly hybridized with
the valence band. Its center of gravity is located at about 3.5 eV
below the top of the valence band. The minority Mn $3d$ states
(lower part) on the contrary, are barely visible on the Figure; they
start to appear at 6 eV. The split band is of mainly O character
with a high Mn contribution. A more close analysis indicates that it
is mainly localized on the 2$p$ orbitals of the nearest O neighbors
of the Mn impurity. As it is visible in the Figure, due to the
isovalent impurity, the Fermi level is located just above the split
band. Holes may be created by doping (either chemically or in the
photoemission process). A partially filled split band corresponds to
an uncompensated oxygen down spin which turns around the localized
Mn up spin. That picture has a great analogy to the Zhang-Rice
singlet (ZRS) state \cite{Zhang88} in cuprates. In cuprates the ZRS
can qualitatively be described by the LSDA+$U$ method in a similar
manner than here.

A localized hole state leads to several consequences. First of all,
it prevents ferromagnetism if the doped holes are all trapped in
localized states. Second, the exciton seen in magneto-optics is
built with holes at the valence band edge and cannot be built with
localized holes. However, as it is visible in Fig.\ \ref{fig2}, the
valence band edge is split in the opposite direction (apparent
ferromagnetic coupling) and to a much smaller amount (about 1/3 of
$\Delta E^v$). Therefore, strictly speaking, magneto-optics does not
measure $N\beta$ but an apparent $N\beta^{app}$ of the opposite sign
and of smaller amplitude. Our LSDA+$U$ calculation for ZnO:Mn
explains this discrepancy between $N \beta^{app}$ measured in
magneto-optics and the pure antiferromagnetic $N \beta$ parameter
(see Table \ref{table2a}). Experimentally, the ferromagnetic sign of
$N \beta^{app}$ was recently unambiguously demonstrated for GaN:Fe
which is not a II-VI SC, however. \cite{Pacuski08} The difference
between $N\beta$ and $N\beta^{app}$ can also be calculated in the
Wigner-Seitz or in other approaches. \cite{Dietl08}

In contrast to ZnO:Mn we find no localization in ZnO:Co, but a
situation quite close to it. In the corresponding DOS (not shown)
the split band has merged with the valence band. It was already
noted that in the LSDA calculations all $3d$ states are much higher
in energy than in the LSDA+$U$ (which contradicts however the
photoemission measurements and is an artefact of LSDA). Therefore,
we find hole localization in LSDA for all compounds besides ZnTe.
Correspondingly, the $|N\beta|$ values are much higher
($N\beta^{MF}=-3.90$, -2.80, -2.43, and -2.00 eV for ZnA:Mn with
A=O, S, Se, and Te; and $N\beta^{MF}=-3.86$, -4.72, -4.30, and -4.25
eV for ZnA:Co). That contradicts the experimental data already in
the weak coupling compounds ZnSe:Mn and ZnTe:Mn. The relevance of
the LSDA+$U$ approach to calculate $N\beta$ for ZnSe:Mn was first
noted in Ref.\ \onlinecite{Sandratskii03} which is in excellent
agreement with our results. On the other hand, the $N\alpha$ values,
are not very much changed by the $U$ parameter.

\section{Discussion}

Before comparing our results with other work let us mention the
limitations of our procedure. After all, the LSDA+$U$ treats
correlation effects only in an approximative manner and neglects
fluctuations. This might explain the discrepancy for ZnO:Co where
state localization is very probable in view of the large difference
between photoemission and magneto-optics. \cite{Pacuski08,Dietl08} A
more sophisticated method to treat correlation effects will probably
refine the picture presented here. It means that the LSDA+$U$
approach underestimates the localization tendency (and probably also
the $|N\beta|$ values) in the strong coupling case. Other error
sources are the limited knowledge on $U$, the use of the effective
mass approximation in the Wigner-Seitz approach, which is
furthermore restricted to only one valence band in difference to the
real band structure.

Our results show good agreement between theory and experiment for
$J_{dd}$, $N\alpha$ and for $N\beta$ in the weak coupling regime
(principally ZnTe and ZnSe). However, in the strong coupling case,
we would like to argue that our calculated $N\beta$ values
correspond neither to the published ones from magneto-optics (see
discussion above) nor to those from photoemission. Since the
photoemission values of -2.7 (-3.4 eV) for ZnO:Mn (ZnO:Co) were
obtained in an indirect way using the perturbation formula of Larson
{\em et al}: \cite{Larson88}
\begin{equation}
N \beta=-\frac{16}{S} t_{pd}^2 \left[
\frac{1}{U_{eff}-\Delta_{eff}}+\frac{1}{\Delta_{eff}}\right] \; .
\label{61}
\end{equation}
The experimental core-level photoemission spectra \cite{Mizokawa02}
were fitted by the configuration interaction (CI) method to a
MnA$_4$ cluster (with the anions A=O, S, Se, and Te) which fixes the
hybridization parameter $t_{pd}$, the Hubbard correlation in the $d$
shell $U_{eff}$ and the effective charge transfer energy between
$p$- and $d$-orbitals $\Delta_{eff}$ (for more details see Ref.\
\onlinecite{Mizokawa02}). The obtained parameters are repeated in
Table \ref{tabs1} and allow to determine $N\beta$ according to Eqn.\
(\ref{61}). The $N\beta$ value of -3.4 eV for ZnO:Co was obtained by
an identical procedure. \cite{Blinowski02} In the same perturbation
approach we may, however, also calculate the nearest-neighbor
exchange: \cite{Larson88}
\begin{equation}
J_{dd}=-\frac{8.8}{2S^2}t_{pd}^4 \left[
\frac{1}{U_{eff}(U_{eff}-\Delta_{eff})^2}-\frac{1}{(\Delta_{eff}-U_{eff})^3}\right]
 \; . \label{62}
\end{equation}
The calculated values are also given in Table \ref{tabs1} and show
large discrepancies to the experimental results (see Table
\ref{table1a} above) especially in the strong coupling case of
ZnO:Mn. Similar discrepancies can be observed by determining the
hybridization parameter $t_{pd}=(pd\sigma)/3-2\sqrt{3}(pd\pi)/9$ by
band structure calculations. \cite{Chanier08} These difficulties
prove that the perturbation formulas (\ref{61},\ref{62}) have a
restricted applicability and have to be treated with care especially
for strong coupling.

\begin{table}
\caption{\label{tabs1} Photoemission data for Hubbard correlation
$U_{eff}$, charge transfer energy $\Delta_{eff}$, and hybridization
parameter $t_{pd}$, as well as the determined $N\beta$ values
according to Ref.\ \onlinecite{Mizokawa02}. The $p$-$d$ and the
nearest-neighbor $d$-$d$ exchange $J_{dd}$ are determined according
to the perturbation formulas of Larson {\em et al}, Ref.\
\onlinecite{Larson88}.}
\begin{ruledtabular}
\begin{tabular}{cccccc}
 & $\Delta_{eff}$(eV) & $U_{eff}$(eV) & $t_{pd}$(eV) & $N\beta$(eV) & $J_{dd}$(meV)
  \\\hline
ZnO:Mn & 7.71 & 9.61 & 0.80 & -2.7 & -25.29\\
ZnS:Mn & 4.21 & 8.41 & 0.65  & -1.3 & -1.29\\
ZnSe:Mn &  3.21 & 8.41 & 0.56  & -1.0 & -0.39 \\
ZnTe:Mn & 2.71 & 8.41 & 0.51  & -0.9 & -0.21 \\
\end{tabular}
\end{ruledtabular}
\end{table}

Being close in spirit to Ref.\ \onlinecite{Dietl08}, our results
deviate nevertheless quite considerably in the numerical values for
$N\beta$ which were assumed there. We found a much smaller coupling
and we believe that the discrepancy with the published photoemission
(PE) values (which are about two times larger than our results)
results from the non-justified use of the perturbative Larson
formula in analyzing the PE data. As a consequence, our magnitude of
the dimensionless coupling parameter $\eta_{\downarrow}=-1.12$ for
ZnO:Mn is much smaller than that estimated in Ref.\
\onlinecite{Dietl08} (between -2.0 and -3.3). It is highly probable,
that the reduced value of $|N\beta|$ will also reduce the proposed
ferromagnetic Curie temperature in ZnO:Co and ZnO:Mn provided that
the doping level is sufficiently high to delocalize the hole states.

The large discrepancies between different experimental and
theoretical approaches for $N\beta$ in the strong coupling regime
point also to the limitations of the oversimplified model
Hamiltonian (\ref{eq41}) in that limit. The $p$-$d$ hybridization
$t_{pd}$ can then no longer be regarded as a perturbation and the
approximation of an infinite valence band width will probably lead
to wrong conclusions. It is highly questionable that the strong
coupling case can still be analyzed in such a manner.

We thank Anatole Stepanov, Sergei Ryabchenko, and Roman Kuzian for
useful discussions. Financial support from the "Dnipro" (14182XB)
program is grateful acknowledged.

\end{document}